\documentclass[aip,apl,onecolumn,superscriptaddress]{revtex4-1}
\usepackage{graphicx}

\begin{document}


\title{Electronic structure of a superconducting topological insulator Sr-doped Bi$_2$Se$_3$} 

\author{C. Q. Han}
\affiliation{Key Laboratory of Artificial Structures and Quantum Control (Ministry of Education), Department of Physics and Astronomy, Shanghai Jiao Tong University, Shanghai 200240, China}
\author{H. Li}
\affiliation{Shanghai Institute of Microsystem and Information Technology, Chinese Academy of Science, Shanghai 200050, China}
\author{W. J. Chen}
\author{Fengfeng Zhu}
\author{Meng-Yu Yao}
\affiliation{Key Laboratory of Artificial Structures and Quantum Control (Ministry of Education), Department of Physics and Astronomy, Shanghai Jiao Tong University, Shanghai 200240, China}
\author{Z. J. Li}
\author{M. Wang}
\author{Bo F. Gao}
\email{bo\_f\_gao@mail.sim.ac.cn}
\affiliation{Shanghai Institute of Microsystem and Information Technology, Chinese Academy of Science, Shanghai 200050, China}
\author{D. D. Guan}
\author{Canhua Liu}
\affiliation{Key Laboratory of Artificial Structures and Quantum Control (Ministry of Education), Department of Physics and Astronomy, Shanghai Jiao Tong University, Shanghai 200240, China}
\affiliation{Collaborative Innovation Center of Advanced Microstructures, Nanjing 210093, China}
\author{C. L. Gao}
\affiliation{Key Laboratory of Artificial Structures and Quantum Control (Ministry of Education), Department of Physics and Astronomy, Shanghai Jiao Tong University, Shanghai 200240, China}
\affiliation{Collaborative Innovation Center of Advanced Microstructures, Nanjing 210093, China}
\affiliation{Department of Physics, Fudan University, Shanghai 200433, China}
\author{Dong Qian}
\email{dqian@sjtu.edu.cn}
\author{Jin-Feng Jia}
\affiliation{Key Laboratory of Artificial Structures and Quantum Control (Ministry of Education), Department of Physics and Astronomy, Shanghai Jiao Tong University, Shanghai 200240, China}
\affiliation{Collaborative Innovation Center of Advanced Microstructures, Nanjing 210093, China}

\date{\today}
\begin{abstract}
Using high-resolution angle-resolved photoemission spectroscopy and scanning tunneling microscopy/spectroscopy, the atomic and low energy electronic structure of the Sr-doped superconducting topological insulators (Sr$_x$Bi$_2$Se$_3$) were studied. Scanning tunneling microscopy shows that most of the Sr atoms are not in the van der Waals gap. After the Sr doping, the Fermi level was found to move a little bit upwards compared with the parent compound Bi$_2$Se$_3$, which is consistent with the low carrier density in this system. The topological surface state was clearly observed and the position of the Dirac point was determined in all doped samples. The surface state is well separated from the bulk conduction bands in the momentum space. The persistence of separated topological surface state combining with small Fermi energy makes this superconducting material a very promising candidate of the time reversal invariant topological superconductor.
\end{abstract}


\maketitle
Topological superconductors (TSCs) are a exotic type of superconductors that attracted extensive research interests in the last several years\cite{TSC1,TSC2,TSC3,TSC4}. TSCs provide a platform to study the exotic quasiparticle named Majorana Fermion (MF) that has been proposed to be potentially useful for the fault-tolerant topological quantum computations\cite{RMP}. Many proposals have been raised to realize the TSC in the real materials\cite{TSC3,FQH,SRO1,SRO2,Wangscience,SC-SOC,SatoPRL,LuPRL}, such as fractional quantum Hall states\cite{FQH}, p-wave superconductors\cite{SRO1,SRO2}, superconductor-topological insulator (TI) heterostructures\cite{TSC3,Wangscience}, s-wave Rashba superconductors\cite{SC-SOC}, spin-orbit coupled nodal superconductors\cite{SatoPRL}, spin-singlet nodal superconductors with magnetic order\cite{LuPRL} and so on. Among those proposals, TSCs based on the TIs have been systematically explored very recently. In order to turn a TI to a superconductor, three kinds of experimental methods have been carried out. The first method is to use the high pressure. Superconductivity in three dimensional (3D) TIs (Bi$_2$Se$_3$, Bi$_2$Te$_3$, Sb$_2$Te$_3$) were achieved under the high pressure\cite{BiSeHP,SbTeHP,BiTeHP}. Unfortunately, it is very difficult to do other experiments under the high pressure circumstance to determine whether they are in TSC state or not. The second method is to use superconducting proximity effect. Superconducting behavior was observed in Bi$_2$Se$_3$ and Bi$_2$Te$_3$ through the proximity effect when they were contacted with a s-wave superconductor\cite{NatureMaterials,Wangscience,XuPRL2014,XuPRL2015}. Some experimental signature of the MFs was recently observed by scanning tunneling microscopy/spectroscopy (STM/STS)\cite{XuPRL2015} in the TI/SC (Bi$_2$Te$_3$/NbSe$_2$) heterostructure. The third method is to dope the bulk TIs. Bulk superconductivity was realized in the doped 3D TIs. Bi$_2$Se$_3$ samples doped with Cu show the superconductivity below $\sim$ 3.8 K\cite{CuBiSe}. However, the impurity phase of Cu$_x$Se in the Cu-doped Bi$_2$Se$_3$ makes this system complicated. The maximum superconducting volume fraction is only about 40\%\cite{CuBiSePRL}. The nature of the superconducting state in the Cu-doped Bi$_2$Se$_3$ is still under debate. Zero bias peak, a signature of the MF, was reported in the point contact experiment\cite{AndoCuBiSe}. However, trivial BCS-type gap function was observed in the very low temperature STS experiment\cite{ZhangtongPRL}.

Very recently, strontium (Sr) doped Bi$_2$Se$_3$ samples also showed bulk superconductivity\cite{SrBiSe,IndiaPRB} with a transition temperature of T$_c$ $\sim$ 2.5 K. The superconducting volume fraction is higher than 80\% at low temperature. Though superconducting Cu-doped and Sr-doped Bi$_2$Se$_3$ are both n-type, the carrier density of Cu-doped sample ($\sim$ 2$\times$10$^{20}$cm$^{-3}$)\cite{CuBiSe} is one order of magnitude higher than Sr-doped sample ($\sim$ 1$\times$10$^{19}$cm$^{-3}$)\cite{SrBiSe,IndiaPRB} with the similar Cu or Sr doping concentration.  The carrier density of Sr-doped Bi$_2$Se$_3$ is close to that of undoped bulk Bi$_2$Se$_3$ crystals\cite{bulkBiSe}. Bi$_2$Se$_3$ has layered crystal structure. The basic unit in Bi$_2$Se$_3$ is the quintuple layer (QL) (Se-Bi-Se-Bi-Se). Between two adjacent QLs, it is van der Waals gap. In Cu-doped Bi$_2$Se$_3$ samples, intercalation of Cu atoms into the van der Waals gap acting as donors was believed to be crucial for the superconductivity\cite{CuBiSe}. STM experiments found lots of Cu atoms on the cleaved surface\cite{CuBiSe}. In Sr-doped samples, Sr atoms were also assumed to locate at the similar position as the Cu-doped samples\cite{SrBiSe}. However, no clearly experimental evidences have been obtained and no electronic structure has been reported on this system. In this work, we investigated the atomic and electronic structures of the superconducting Sr-doped Bi$_2$Se$_3$ samples with different Sr concentration using STM/STS and angle-resolved photoemission spectroscopy (ARPES). Our results do not support the scenario of Sr intercalation into the van der Waals gap. After Sr doping, the Fermi level moves upwards only several tens of meV. The topological surface state remains and well separated from the bulk bands. Sr-doped Bi$_2$Se$_3$ is better than Cu-doped Bi$_2$Se$_3$ for TSC studies.

Single crystals of Sr$_x$Bi$_2$Se$_3$ were grown by melting stoichiometric mixtures of high purity elements Bi (99.999\%), Sr (99.95\%), and Se (99.999\%) in sealed evacuated quartz tubes at 850$^{\circ}$ for 48 h, followed by a slow cooling to 610$^{\circ}$ at a rate of 3$^{\circ}$/h. Then, the quartz tubes were taken out and the samples were quenched down into iced water. The concentration of Sr in the crystals was determined by
inductively coupled plasma (ICP) mass spectrometry. ARPES experiments were performed using 50 $\sim$ 120 eV photons at Advanced Light Source beamline 4.0.3 using Scienta R4000 analyzer with base pressures better than 5$\times$10$^{-11}$ Torr at 30K. Energy resolution is better than 15 meV and angular resolution is better than 1\% of the surface Brillouin zone. DC magnetic susceptibility was measured on a Quantum Design Physical Property Measurement System (PPMS). STM/STS experiments were carried out at 4.2 K and 1 K.

\begin{figure}
\includegraphics[width=8.5cm]{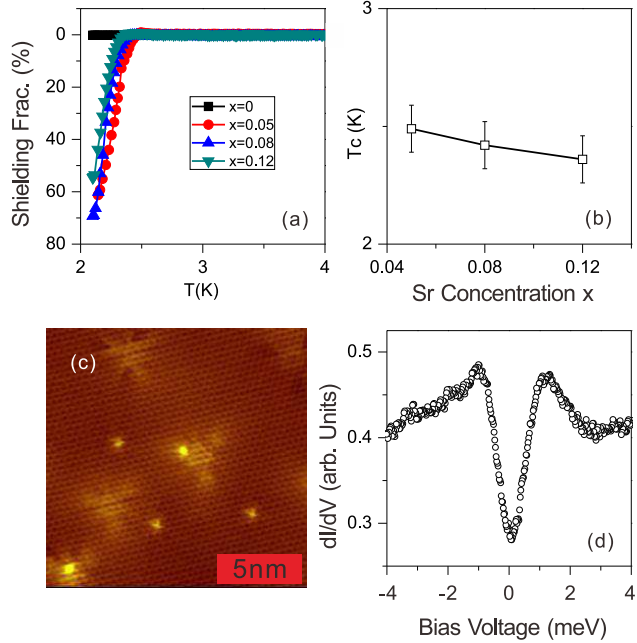}
\caption{(a) DC Magnetization measurements for the Sr-doped Bi$_2$Se$_3$. The Sr concentration x is x = 0, 0.05, 0.08 and 0.12. All doped samples exhibit the superconducting transition. (b) Tc as a function as Sr concentration. (c) STM surface morphology of the cleaved surface of x=0.08 samples. Tunneling current I=252 pA. Bias voltage V=5.7 mV. (d) Superconducting gap on the surface of (c) measured at 1K.}
\end{figure}

Before the spectroscopy experiments, the bulk superconductivity were confirmed by DC magnetic susceptibility measurements. Figure 1(a) shows the temperature dependence of the DC magnetizations of the Sr$_x$Bi$_2$Se$_3$ samples with x = 0, 0.05, 0.08, 0.12 under 20 Guass external magnetic field along the c-axis. Except the parent samples (Bi$_2$Se$_3$), all doped samples show clear diamagnetic signals at low temperature. At 2.1 K, the superconducting shielding fraction is $\sim$ 60\%. Figure 1(b) shows the superconducting temperature (Tc) as a function of the Sr doping concentration. Tc is nearly constant. Figure 1(c) shows a typical STM surface morphology image of the cleaved surface of x = 0.08 samples. It is well known that Bi$_2$Se$_3$ samples break at the van der Waals gap when cleaved. The top surface is Se-terminated. In Fig. 1(c), hexagonal lattice is resolvable in STM. Though there are several defects, the surface is much cleaner than that of the Cu-doped Bi$_2$Se$_3$. Lots of intercalated Cu atoms were observed in Cu-doped Bi$_2$Se$_3$\cite{CuBiSe}, but on our Sr-doped samples, the surface morphology is like undoped Bi$_2$Se$_3$. We conclude that most of the doped Sr atoms are not in the van der Waals gap. They should be within the Se-Bi-Se-Bi-Se QL. So far, we can not give more information about the exact position of Sr atom within the QL by STM. High-resolution TEM experiments are needed. Superconducting gap was observed everywhere on the cleaved surface as Fig. 1(c). Figure 1(d) presents the superconducting STS near the Fermi level taken at 1K. Well defined superconducting coherence peaks were observed. It is worth noting that the conductance (dI/dV) at zero bias voltage is not zero in the STS curve. Though we can not 100\% exclude possible exotic origin of none zero conductance at zero bias voltage, we think it is most likely caused by the temperature effect. Our measurement temperature is too high compared with low T$c$. STS experiments under the order of $\sim$ 100 mK may provide critical information about the nature of superconducting state in Sr-doped samples in the future.

\begin{figure}
\includegraphics[width=8cm]{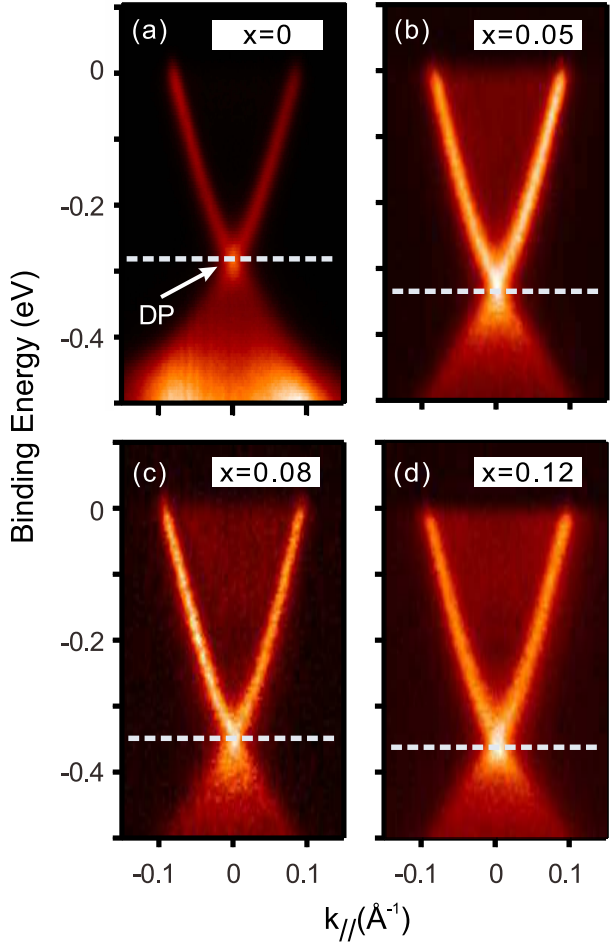}
\caption{Low energy ARPES spectra of the Sr$_x$Bi$_2$Se$_3$ sample: (a) x=0 (b) x=0.05 (c) x=0.08 (d) x=0.12.``DP" labels the Dirac point. The white dashed lines indicate the position of the Dirac point. The incident photon energy is 71 eV.}
\end{figure}

Figure 2 presents the low energy ARPES spectra of Sr$_x$Bi$_2$Se$_3$. In order to emphasize the surface states, we choose the suitable incident photon energy (h$\nu$=71eV) to suppress the signals of the bulk conduction bands. For parent compound Bi$_2$Se$_3$ (Fig. 2(a)), the Dirac point is at $\sim$ 290 meV below the Fermi level. After Sr doping, very clearly, the Dirac cone like surface states remain in all doped samples. Sr doping does not destroy the topological properties. On the other hand, compared with the parent compound Bi$_2$Se$_3$, the Dirac point moves to higher binding energy with Sr doping. In x=0.05 samples (Fig. 2(b)), the Dirac point locates at the binding energy of $\sim$ 320 meV below the Fermi level and there is no obvious change on the Fermi velocity of the surface state. In another word, the topological surface state move downwards by $\sim$ 30 meV rigidly within our experimental uncertainty after x=0.05 Sr doping. Sr provides weak electron doping. In Cu-doped Bi$_2$Se$_3$ samples, the Dirac point was found to be at about 450 meV below the Fermi level\cite{CuBiSeNatPhy}, which means Cu atoms provide much more electrons to the materials. The slightly downward movement of the surface state (or upward movement of the Fermi level) is consistent with the bulk carrier density measurement as reported\cite{SrBiSe,IndiaPRB}. With more Sr doping, the Dirac cone keeps moving downwards. The Dirac point locates at $\sim$ 340 meV in x=0.08 sample and at $\sim$ 360 meV in x = 0.12 below the Fermi level.

\begin{figure}
\includegraphics[width=8.5cm]{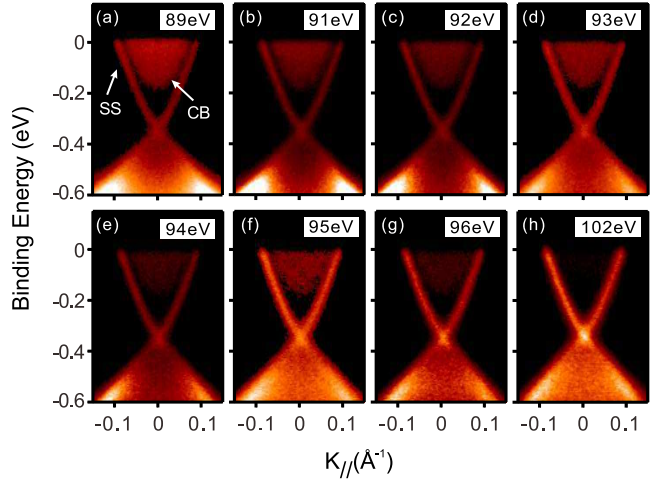}
\caption{Low energy ARPES spectra of Sr$_{0.08}$Bi$_2$Te$_3$ along the $K- \Gamma - K$ direction using different incident photon energy: h$\nu$ = (a) 89 eV, (b) 91 eV, (c) 92 eV, (d) 93 eV, (e) 94 eV, (f) 95 eV, (g) 96 eV, (h) 102 eV.``SS" lables the surface state. ``CB" labels the bulk conduction bands}
\end{figure}

To be a TSC, the exact position of the surface state and the bulk states of a superconducting doped TI is crucial\cite{FuPRLCuBiSe, PRLTeo, PRLHosur}. We carefully looked at the bulk conduction bands in x=0.08 samples by tuning the incident photon energy (changing k$z$). Figure 3 presents the low energy ARPES spectra using different incident photon energies. Different from the spectra using 71 eV photon in Fig. 2, both surface states and the bulk conduction bands are observed. Ignoring the variation of intensity, the dispersion of the surface states does not change when changing the incident photon energy. Only the dispersion of the bulk conduction bands changes. In Fig. 3 (a), the bottom of the bulk conduction band reaches the minimum at h$\nu$ = 89 eV. For all photon energy (all k$_z$), it is clear that the surface states and the bulk conduction bands are well separated in the momentum space. In theory, the superconducting doped TI would be topological nontrivial if it has suitable bulk band structure and persists topological surface state (surface Dirac cone)\cite{FuPRLCuBiSe, PRLTeo, PRLHosur}. Fu et al.\cite{FuPRLCuBiSe} proposed that the superconducting doped TI can be a 3D odd parity TSC depending on the $m^*/\mu$ and U/V if its bulk Fermi surface under non-superconducting state encloses an odd number of time-reversal invariant momenta (in other word, topological surface state persists after doping). m$^*$ is the effective mass or half of the band gap of a Dirac-like model (for Bi$_2$Se$_3$ m* $\sim$ 0.15 eV). $\mu$ is chemical potential (energy different between the Fermi level and the Dirac point). U and V are the intraorbital and interorbital interactions, respectively. For Cu-doped Bi$_2$Se$_3$, $m^*/\mu$ was estimated as 1/3 ($\mu$ $\sim$ 0.4 eV)\cite{CuBiSeNatPhy}. For Sr-doped Bi$_2$Se$_3$, m$^*$ remains the same and $\mu$ is about 0.34 eV for x=0.08. So $m^*/\mu \sim$ 0.44 that is still small enough to realize the rare odd parity 3D TSC state according to the phase diagram proposed in Ref. 27 though U/V is difficult to estimate. A 3D TSC is fully gapped in the bulk but have gapless surface Andreev bound states\cite{FuPRLCuBiSe} that can be checked by very low temperature STM in the future. On the other hand, if it is not odd parity 3D TSC, the topological surface state could also form nontrivial 2D TSC if the surface state is separated from the bulk states in momentum\cite{PRLTeo, PRLHosur}. The 2D TSC state on the surface is fully gapped but supports zero-energy MFs localized in the vortex core. Theoretical model calculations suggest that the number of MFs at the end of a vortex line is partly derived from the bulk states\cite{PRLTeo, PRLHosur}. If the Fermi level is inside the bulk gap or not very far away from the bulk conduction band minimum (CBM) of a TI, then an odd number of MFs (e.g., one) is expected be exist at the end of a vortex line. If the Fermi energy is far above CBM, the end of each vortex line will have an even number of MFs (e.g., zero or two), which is nearly equal to the trivial superconductor. Hosur et al.\cite{PRLHosur} have predicted that transition will occur when the Fermi level is higher than 0.24 eV above CBM. For Cu-doped Bi$_2$Se$_3$, Fermi level is about 0.25 eV above CBM, which is very close to the transition point. For Sr-doped Bi$_2$Se$_3$, Fermi level is less than 0.2 eV above CBM ($\sim$ 0.18 eV for x=0.08). It is within the nontrival region. Therefore, Sr-doped samples have large possibility to possess the useful 2D TSC state on the surface.

At last, we gave some discussion on the possible position of Sr atoms. Naively speaking, in the crystal, Sr can be possibly locate at four different positions: i)intercalate into the van der Waals gap; ii)intercalate into the QL of Bi$_2$Se$_3$. iii) substitute Bi; iv) substitute Se. Considering the chemical activity of the Sr, substitution of Se is very unlikely. Intercalation of the Sr atoms will provide electrons to the Fermi sea. Substitution of Bi will provide holes to the Fermi sea. Since we hardly find the intercalated Sr atoms on the surface, will most of the Sr substite Bi? The answer is no. Because if most of the Sr atoms substitute Bi atoms like Ca doping does\cite{Cadoping}, Sr-doped samples should be p-type instead of n-type. In order to realize low carrier density, we propose the coexistence of the substitution of Bi and the intercalation of Sr atoms into the QL. Our proposal can be examined by high resolution TEM in the future.

In summary, we have performed ARPES and STM/STS studies on the superconducting doped TI Sr$_x$Bi$_2$Se$_3$. We found that most of the Sr atoms are not in the van der Waals gap. Superconducting gap was observed on the clean surface. With Sr doping, the energy bands rigidly shift to the higher binding energy by several tens of meV. The Fermi level locates less than 0.2 eV above the CBM. On all doped samples, topological surface state remains and well separated from the bulk bands. The separated topological surface state together with small Fermi energy that satisfy the theoretical requirements of TSC imply that this superconductor is a promising candidate for TSC researches.

This work is supported by National Basic Research Program of China (Grants No. 2012CB927401, No. 2013CB921902), National Natural Science Foundation of China (Grants No. 11574201, No. 11521404, No. 11134008, No. 11174199, No. 11374206, No. 11274228, No. 11227404, No. 91421312, and No. 91221302), Shanghai Committee of Science and Technology (No. 12JC140530), C.L.G. acknowledges support from the Shu Guang project, which is supported by the Shanghai Municipal Education Commission and Shanghai Education Development Foundation. D.Q. acknowledges support from the Top-notch Young Talents Program and the Program for Professor of Special Appointment (Eastern Scholar). The Advanced Light Source is supported by the Director, Office of Science, Office of Basic Energy Sciences, of the US Department of Energy under Contract No. DE-AC02-05CH11231.

---------------------------------------------------------------

\end{document}